# Reconstruction of Electrical Impedance Tomography Using Fish School Search, Non-Blind Search and Genetic Algorithm

Valter A. F. Barbosa [a,*], Reiga R. Ribeiro [a], Allan R. S. Feitosa [a],
Victor L. B. A. Silva [b], Arthur D. D. Rocha [b], Rafaela C. Freitas [b],
Ricardo E. Souza [a], Wellington P. Santos [a,b]

[a]Departamento de Engenharia Biomédica, Universidade Federal de Pernambuco, Cidade Universitária, Recife, PE, 50670-901, Brazil
[b]Escola Politécnica da Universidade de Pernambuco, POLI-UPE, Madalena, Recife, PE, 50720-001, Brazil

**ABSTRACT**

Electrical Impedance Tomography (EIT) is a noninvasive imaging technique that does not use ionizing radiation, with application both in environmental sciences and in health. Image reconstruction is performed by solving an inverse problem and ill-posed. Evolutionary Computation and Swarm Intelligence have become a source of methods for solving inverse problems. Fish School Search (FSS) is a promising search and optimization method, based on the dynamics of schools of fish. In this article we present a method for reconstruction of EIT images based on FSS and Non-Blind Search (NBS). The method was evaluated using numerical phantoms consisting of electrical conductivity images with subjects in the center, between the center and the edge and on the edge of a circular section, with meshes of 415 finite elements. We performed 20 simulations for each configuration. Results showed that both FSS and FSS-NBS were able to converge faster than genetic algorithms.

## INTRODUCTION

Ionizing radiation is commonly used in medical image machines, as mammography, positron emission tomography or x-rays. Besides the benefits that using those electromagnetic waves may provide, there are many associated risks to whom operates those machines or is submitted to these kind of exams. Also, the prolonged exposition to ionizing radiation may cause many diseases, such as cancer (Rolnik & Seleghim Jr, 2006). Possibly, this issue is one of the most debated subjects in Public Health all over the world, strengthening the search for imaging technologies that are: efficient, low-cost, simple and safe to those that uses them.

A promising imaging technique, that does not use ionizing radiation, is Electric Impedance

Tomography (EIT) (Bera & Nagaraju, 2014; Rolnik & Seleghim Jr, 2006). EIT is about a non-invasive technique that builds images of an interior body (or any object), using electrical properties, measured over the surface of interest. Those measurements are acquired from electrodes' disposition around the transversal section of interest, and the application of a low amplitude and high frequency current through them creates an electric potential, known as "border potential". This low-voltage signal is measured and, in a computer, they are used in a reconstruction algorithm that rebuilds the image of the body's inside region of interest (Rasteiro, Silva, Garcia, & Faia, 2011; Tehrani, Jin, McEwan, & van Schaik, 2010; Brown, Barber, & Seagar, 1985).

In medical sciences, EIT can be applied in several situations, such as: breast cancer (Cherepenin et al., 2001), pulmonary ventilation monitoring (Alves, Amato, Terra,Vargas, & Caruso, 2014), in the detection of pulmonary embolism or blood clots in the lungs (Cheney, Isaacson, & Newell, 1999). Likewise, it can be applied in fields as Botanics, generating images of the trees' trunks' insides, allowing the knowledge of its biological conditions without damaging it (Filipowicz & Rymarczyk, 2012); in monitoring multiphasic outflow in pipes (Rolnik & Seleghim Jr, 2006); in Geophysics, EIT is largely used to find underground storage of mineral and different geological formations (Cheney et al., 1999).

When compared with techniques, like Magnetic Resonance Tomography, or X-Ray Tomography, EIT has a relatively low cost, since, in simple manners, it needs an equipment able to generate and measure current and electric potential, and a computer, able to rebuild the image (Tehrani et al., 2010). Also, since it uses only the electrical properties (conductivity and permittivity) of the body, there are no associated risk to its use, unlike acquisition methods that uses ionizing radiation.

However, Electric Impedance Tomography images have, still, low resolution and undefined borders, which harms its popularity and diffusion among the imaging field. This motivates researchers of EIT to seek new methods of image reconstruction that are also able to overcome these techniques disabilities, creating images with good resolution and low computational cost, making of it a reliable and easy tool on diseases' diagnostics.

Mathematically, EIT reconstruction problem is known as ill-posed and ill-conditioned, meaning that there are not only one solution (image) for a given potential border distribution. Many algorithms are applied in order to solve EIT problem, and, however, the image generated is not totally reliable or well defined (Rolnik & Seleghim Jr, 2006).

Thus, an alternative way used in the attempt of solving the EIT problem is managing it as an optimization problem, which the objective is minimize the relative error between the measured border potential of an object and the calculated border potential of the solution candidate (Feitosa, Ribeiro, Barbosa, de Souza, & dos Santos, 2014; Ribeiro, Feitosa, de Souza, & dos Santos, 2014a, 2014b, 2014c).

A heuristic that may be used in order to solve this as an optimization problem is the Fish School Search (FSS) (Bastos-Filho, de Lima Neto, Lins, Nascimento, & Lima, 2008; Bastos-Filho & Guimarães, 2015). This technique is inspired in fish schools' behavior on food search. The search process on FSS is made by a population which its individuals (the fishes) has a limited memory. Each school represents a possible solution for the system. The fishes interact among each other and with the environment that surrounds them, and, by influence of the collective and individual movement's operator and food operator, the school increases the possibility of convergence to the food surroundings, which means the best position and solution to that problem (Lins, Bastos-Filho, Nascimento, Junior, & de Lima-Neto, 2012).

In this work, a relatively simple approach to image reconstruction problem of EIT is proposed, using Fish School Search (FSS). However, it was modified, presenting two ways of solution candidates (fish) initialization: one completely random and other, among the random candidates, one solution derived from the Gauss-Newton reconstruction method. Taking into account Saha and Bandyopadhyay (2008) this initialization method was called Non-Blind Search.

This work is organized as following. In section Materials and Methods we present a brief on the theoretical foundations of Electrical Impedance Tomography and inverse problems, Fish School Search, Non-Blind Search, the experimental infrastructure, and our proposal. In section Results and Discussion we present experimental results and detailed discussion. Finally, in section Conclusion we present conclusions and some highlights of future developments.

## MATERIALS AND METHODS

### Electrical Impedance Tomography: Mathematical Formulation And Reconstruction Problems

In EIT the estimate of the electrical conductivity distribution, inside a heterogeneous body or object, is made by the resolution of a partial differential equation named Poisson's Equation (Borcea, 2002; Cheney et al., 1999). The process to obtain the Poisson's Equation is originated from the Maxwell's Equations, it is starting from the Gauss's law in point form (Tombe, 2012):

$$\nabla \cdot \vec{D} = \rho \qquad (1)$$

Where $\nabla \cdot$ is the divergent operator, $\rho$ is the free electric charge in the interest region, and $\vec{D}$ is the electric elasticity given by the multiplication of the electrical conductive distribution $\sigma(\vec{u})$ in the point $\vec{u} = (x, y, z)$ and the Electrical field $\vec{E}$, as a follow:

$$\vec{D} = \sigma(\vec{u})\vec{E} \qquad (2)$$

Knowing that the electrical field $\vec{E}$ is determined by the negative gradient (denoted by the nabla symbol - $\nabla$) of the electrical potentials ($\phi(\vec{u})$), we have that:

$$\vec{E} = -\nabla \phi(\vec{u}) \qquad (3)$$

In the reconstruction problem of EIT images we consider that there is no free electric charge in the interest region (i.e. $\rho = 0$). Taking that into account and replacing the Equations (2) and (3) in (1) we get the Poisson's Equation (Borcea, 2002; Cheney et al., 1999) as given below:

$$-\nabla \cdot [\sigma(\vec{u})\nabla \phi(\vec{u})] = 0 \qquad (4)$$

Besides, we also need to consider the following boundary conditions (Borcea, 2002):

$$\phi_{ext}(\vec{u}) = \phi(\vec{u}), \forall \vec{u} \in \partial\Omega \tag{5}$$

$$I(\vec{u}) = -\sigma(\vec{u})\nabla\phi(\vec{u}) \cdot \hat{n}(\vec{u}), \forall \vec{u} \in \partial\Omega \tag{6}$$

Where $\vec{u} = (x, y, z)$ is the position of a given object, $\phi(\vec{u})$ is the potentials' global distribution, $\phi_{ext}(\vec{u})$ is the electric potentials distribution on the surface electrodes, $I(\vec{u})$ is the electric current applied on the interest region's surface, $\sigma(\vec{u})$ is the electric conductivity distribution (i.e., the goal image), $\Omega$ is the interest volume, $\partial\Omega$ is the volume border and $\hat{n}(\vec{u})$ is the border's normal vector on $\vec{u} \in \partial\Omega$ position.

Finding the electric potential of the surface electrodes $\phi_{ext}(\vec{u})$, given the electric currents $I(\vec{u})$ and the conductivity distribution $\sigma(\vec{u})$ is named EIT's Direct Problem, and modeled by the following relation:

$$\phi_{ext}(\vec{u}) = f(I(\vec{u}), \sigma(\vec{u})), \forall \vec{u} \in \partial\Omega \wedge \vec{u} \in \Omega \tag{7}$$

In Direct Problem's situation, the surface electric potentials estimative, when the internal conductivity distribution is already known, is calculated using the Poisson's equation, shown in (4). Considering the contour condition, given by the following equation:

$$\sigma\frac{\partial\phi}{\partial\hat{n}} = J \tag{8}$$

Where $\hat{n}$ is the surface's normal vector and $J$ corresponds to the electric current density (Baker, 1989). It is important to emphasize that there are no analytical solutions to (4) and (8), for an arbitrary given domain $\Omega$.

Nevertheless, an approximate solution to the border's potentials may be obtained by the Finite Elements Method (FEM), which converts the nonlinear system in (4) and (8) in the following linear equation's system (Bathe, 2006; Castro Martins, Camargo, Lima, Amato, & Tsuzuki, 2012):

$$K(\sigma) \cdot \Phi - C = 0 \tag{9}$$

Where $K(\sigma)$ is a conductivity-dependent ($\sigma$) coefficients matrix and $C$ is a constant's values vector. In this way, it is possible to obtain an approximated value for the border potentials $\Phi$, known as conductivity distribution $\sigma$.

While the conductivity distribution determination problem $\sigma(\vec{u})$ (tomographic image), given $I(\vec{u})$ and $\phi_{ext}(\vec{u})$ is known as EIT Inverse Problem, modeled as follows:

$$\sigma(\vec{u}) = f^{-1}(I(\vec{u}), \phi_{ext}(\vec{u})), \forall \vec{u} \in \partial\Omega \wedge \vec{u} \in \Omega \tag{10}$$

In this situation it is possible to obtain the conductivity distribution $\sigma(\vec{u})$ by Poisson's equation solution (4), considering the contour conditions, mentioned in Equations (5) and (6).

## The Objective Function On EIT's Reconstruction Images

To consider the EIT's image reconstruction as an optimization problem, the relative squared error was considered as the objective function (fitness function) between the object's border measured electric potentials and the calculated ones, originated by the generated images, given by the candidate search algorithm (Feitosa et al., 2014; Ribeiro et al., 2014a, 2014b, 2014c). In Equation (11), the fitness function ($f_o(x)$) is given by:

$$f_o(x) = \left( \frac{\sum_{i=1}^{n_e} (U_i(x) - V_i)^2}{\sum_{i=1}^{n_e} (V_i)^2} \right)^{\frac{1}{2}} \quad (11)$$

$$V = (V_1, V_2, ..., V_{n_e})^T \quad (12)$$

$$U(x) = (U_1(x), U_2(x), ..., U_{n_e}(x))^T \quad (13)$$

Where $x$ represents a solution candidate on the search algorithm, $V$ and $U(x)$ the electrical conductivity distribution, measured and calculated on the border, and, $n_e$ the border's electrodes number.

## Fish School Search

Fish School Search (FSS) algorithm is a meta-heuristic based on fish behavior for food search, developed by Bastos Filho e Lima Neto, in 2007 (Bastos-Filho et al., 2008; Bastos-Filho & Guimarães, 2015). The search process on FSS is made by a population which its individuals (the fishes) have limited memory. Also, each fish in the school represents a point on fitness function domain. The FSS algorithm has four operators that can be classified in two classes: food and swimming.

### Food Operator

Aiming to find more food, the fish on the school may move. Therefore, accordingly to its positions, each fish can be heavier or lighter (increase or decrease its weight), depending on how close they are from food (Lins et al., 2012). The food operator, then, quantifies how successful a fish is, due its fitness function variation. The fish weight is given by Equation 14, below:

$$W_i(t+1) = W_i(t) + \frac{f[x_i(t+1)] - f[x_i(t)]}{\max\{|f[x_i(t+1)] - f[x_i(t)]|\}} \quad (14)$$

Where $W_i(t)$, $f[x_i(t)]$ represents the fish '$i$' weight and its fitness function value at $x_i(t)$, respectively. According to Bastos-Filho et al (2008) the concept of food is related to the fitness function, i.e., in a minimization problem the amount of food in a region is inversely proportional to the function evaluation in this region. Thus, in this case, the fish weight is given by the following expression: $W_i(t+1) = W_i(t) + \frac{f[x_i(t)] - f[x_i(t+1)]}{\max\{|f[x_i(t)] - f[x_i(t+1)]|\}}$.

*Swimming Operators*

The swimming operators are responsible for the fish movements when they are in the food search, and are named as: individual movement operator, collective-instinctive movement operator and collective-volitive movement operator, explained in details as below.

The first swimming operator is the individual movement executed at the beginning of each algorithm's iteration, where each fish is displaced to a random position of its surroundings. An important characteristic of this movement is that the fish only executes the individual movement if the new position, randomly determined, is better than the previous one, meaning that it only occurs if the new position provides a better fitness function value. Otherwise, the fish will not execute the movement.

The individual movement of each fish is given in Equation (15), $rand[-1,1]$ is a vector composed by several numbers randomly generated with values between $[-1,1]$, and $step_{ind}$ is a parameter that represents the fish ability of exploration on the individual movement. After the individual movement's calculus, the fish position is updated by Equation (16).

$$\Delta x_{ind_i}(t+1) = step_{ind} \cdot rand[-1,1] \qquad (15)$$
$$x_{ind_i}(t+1) = x_{ind_i}(t) + \Delta x_{ind_i}(t+1) \qquad (16)$$

This movement can be understood as a disturbance in the fish position, to guarantee a wider way to explore the search space. Therefore, to assure convergence at the end of the algorithm's operation, the value of $step_{ind}$ linearly decays, accordingly to Equation (17), where $step_{ind_{init}}$ and $step_{ind_{end}}$ are the initial and final values of $step_{ind}$, and, *iterations* is the maximum iterations possible value of the algorithm.

$$step_{ind}(t+1) = step_{ind}(t+1) - \frac{step_{ind_{init}} - step_{ind_{end}}}{iterations} \qquad (17)$$

The second swimming operator of the FSS is the collective-instinctive movement. Is the one where the most well succeeded fishes on their individual movements attracts to themselves other fishes. To execute this movement, it is considered the resultant direction vector, $I(t)$, given by the weighted average of all individual movements of each fish, having as weight, its fitness value variation, given in Equation (18), where $N$ is the total of fishes in the school. In the same way of the feeding operator, in minimization problems the fitness variation in Equation (18) must be inverted. After the direction vector calculation, the fish position is updated, as shown in Equation (19).

$$I(t) = \frac{\sum_{i=1}^{N} \Delta x_{ind_i} f[x_i(t+1)] - f[x_i(t)]}{\sum_{i=1}^{N} f[x_i(t+1)] - f[x_i(t)]} \quad (18)$$

$$x_i(t+1) = x_i(t) + I(t) \quad (19)$$

The collective-volitive movement (the third and the last swimming operator) is based on the school's global performance (Lins et al., 2012). The collective-volitive movement is the tool that provides to the algorithm the ability to adjust the search space radius. Therefore, if the fish global weight increases, the search is characterized as well-succeeded and the fish radius search must diminish; otherwise, the same given search radius must increase, in order to enlarge the fish exploration, aiming to find better regions. In this movement, the fish's position is updated in relation to the school's mass center, as showed in (20).

$$Bary(t) = \frac{\sum_{i=1}^{N} x_i(t) W_i(t)}{\sum_{i=1}^{N} W_i(t)} \quad (20)$$

Still, each fish's movement is made by (21), if the school's weight is increasing, or by Equation (21), if the school's weight is decreasing. Also, in the same equations, mentioned above, $rand[0,1]$ is a vector which values are randomly generated between $[0,1]$, and $step_{vol}$ is the parameter that represents the intensity of the fish search adjust intensity.

$$x(t+1) = x(t) - step_{vol} \cdot rand[0,1](x(t) - Bary(t)) \quad (21)$$

$$x(t+1) = x(t) + step_{vol} \cdot rand[0,1](x(t) - Bary(t)) \quad (22)$$

Fish School Search algorithm's pseudocode is given in Algorithm 1.

**Algorithm 1:** Fish School Search
1. Initialize all the fish in random positions
2. Repeat the following (a) to (f) until some stopping criterion is met
    a) For each fish do:
        i) Execute the individual movement
        ii) Evaluate the fitness function
        iii) Execute the feeding operator
    b) Calculate the resulting direction vector - I(t).
    c) For each fish do:
        i) Execute the collective-instinctive movement
    d) Calculate the barycenter.
    e) For each fish do:
        i) Execute the collective-volitive movement
    f) Update the values of individual and collective-volitive step
3. Select the fish in the final school that has better fitness.

## Genetic Algorithm

Genetic Algorithm (GA) consists in a heuristic iterative process applied in search and optimization problems constituted by metaphors inspired by the Evolutions Theories and Genetic principles (Eberhart & Shi, 2011). The GA pseudocode is given in Algorithm 2.

**Algorithm 2:** Genetic Algorithm
1. Initialize a random initial population
2. Repeat the following (a) to (e) until the stopping criterion is met
    a) Evaluate the fitness function to each individual
    b) Parent selection: Using Roulette Wheel individuals are selected to be recombined
    c) Recombination: New individuals are generated through 2-points crossover
    d) Mutation: gene of descendants is randomly selected and modified.
    e) Survivor selection: individuals of the next generation are selected using elitism and roulette wheel.
3. Select the individual's final population that has better fitness

## Non-Blind Search

According to Saha and Bandyopadhyay (2008) in order to avoid a totally random search and accelerate the optimization algorithm convergence, we need define the initial population of candidate solutions using solutions obtained by imprecise, simple and direct methods (Saha & Bandyopadhyay, 2008). Our hypothesis is that the using of the FSS to solve the ill-posed problem of EIT can get reasonable solutions using a small number of iterations, when the initial population has one candidate solution constructed using noisy versions of the solution obtained by the Gauss-Newton method.

## Electrical Impedance Tomography and Diffuse Optical Tomography Reconstruction Software

Electrical Impedance Tomography and Diffuse Optical Tomography Reconstruction Software (EIDORS) is an open source software developed for MATLAB/Octave that has as goal to solve the direct and inverse problems of the electrical impedance tomography and diffuse optical tomography (Adler & Lionheart, 2006; Vauhkonen, Lionheart, Heikkinen, Vauhkonen, & Kaipio, 2001). This software allows its free modification, thus, we can easily adapt it to the problem of this work. With EIDORS it is possible simulate different kinds of meshes of finite elements that represents computationally one cross-section of an object as well as its internal conductivity distribution in the form of colors.

## Proposed Method And Experiments

Using EIDORS, three ground-truth images were created with mesh of 415 finite elements. The goal was detecting irregular objects isolated in three positions: in the center, between the center and the edge and on the edge of the circular domain. The EIDORS parameters to create these images were: 16 electrodes, two-dimensional mesh (2D) with elements density 'b' and electrode refinement level '2'. The Figure 1 shows the three ground-truth images considered in this work.

Twenty (20) simulations for each ground-truth image using Fish School Search without and with Non-Blind Search and Genetic Algorithm were performed. The relative squared error was used between the distribution of electrical potentials measured and calculated at the edge as fitness function for the heuristics considered in this work. Solution candidates (fish to FSS and individuals to GA) are real-valued vectors utilized as theoretical abstractions for possible distributions conductivity, where each dimension of the vector corresponds to a particular finite element on the mesh.

For the simulations using Fish School Search, 100 fishes (solution candidates) were set as the school's population, and the following parameters were defined: $W_0 = 100$, $step_{ind_{init}} = 0.01$, $step_{ind_{end}} = 0.0001$ and $step_{vol} = 2 step_{ind}$. Whereas for the genetic algorithm we used a population with 100 individuals, selection for the 10 best evaluated individuals, probability of recombination and mutation in 100 % and elitism of 10 individuals. The stop criterion for all methods was the number of iterations in 500 iterations.

*Figure 1. Ground-truth images with 415 elements for the object placed in (a) the center, (b) between the center and the edge and (c) on the edge of the circular domain.*

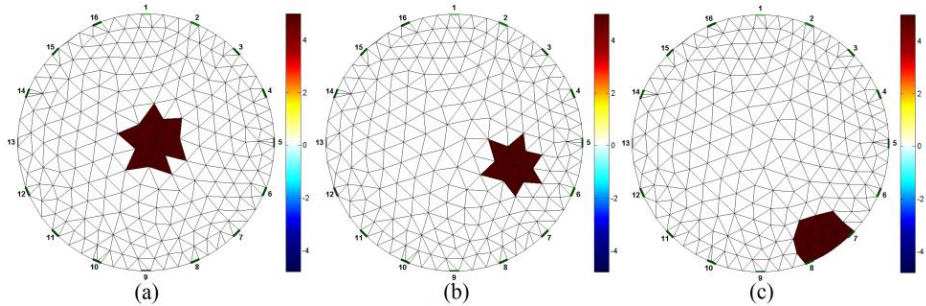

(a)     (b)     (c)

## RESULTS AND DISCUSSION

In this section, the obtained results, generated by FSS with and without non-blind search will be compared with the previous results, obtained by the EIT research group of UFPE, using genetic algorithm. The main reason of this choice is based on the fact that genetic algorithm produces very good and reliable results, making of it a comparison parameter for the new algorithms, used in this work.

The Table 1 gives the fitness values of the best and the worst solutions obtained in twenty simulations for each ground-truth image and each method. It is also given the mean and the standard deviation of the results obtained. As a minimization problem, it is worth noting that the best solutions is given by the lower values. Analyzing this table, we can notice that the perform of the three methods are similar. Actually, neither method outperformed the others. One can say that in some case one method gave the best solution, for example, in the case where the object is between the center and the edge FSS+NBS gave the best solution, but the best solution for object in the edge was obtained by GA. In the same way, between the worst solutions found by the methods, FSS gave the best and obtained the smaller standard deviation for object in the center.

Table 1. The best and worst solutions, the mean and standard deviation for 20 simulations for FSS, FSS+NBS and GA. The results in C, CE and E are for the object in center, between the center and the edge and on the edge of the circular domain, respectively.

|  |  | Best | Worst | Mean | Stnd. deviation |
|---|---|---|---|---|---|
| FSS | C | 0.0198 | 0.0242 | 0.0219 | 0.0012 |
|  | CE | 0.0245 | 0.0306 | 0.0265 | 0.0013 |
|  | E | 0.0242 | 0.0600 | 0.0351 | 0.0089 |
| FSS+NBS | C | 0.0148 | 0.0308 | 0.0186 | 0.0038 |
|  | CE | 0.0174 | 0.0286 | 0.0229 | 0.0032 |
|  | E | 0.0376 | 0.0590 | 0.0446 | 0.0054 |
| GA | C | 0.0182 | 0.0279 | 0.0220 | 0.0027 |
|  | CE | 0.0176 | 0.0276 | 0.0223 | 0.0029 |
|  | E | 0.0208 | 0.0467 | 0.0338 | 0.0069 |

The reconstruction algorithm's behavior can be investigated through results' visual analysis, obtained from the reconstruction images', generated by the algorithms discussed in this article. Figures 2, 3 and 4 show the reconstruction results acquired from the Fish School Search algorithms without (FSS) and with (FSS-NBS) non-blind search, and Genetic Algorithm, respectively, for objects placed in the center (a1, a2 and a3), between the center and the edge (b1, b2 and b3) and on the edge (c1, c2 and c3); in the circular domain for 50, 300 and 500 iterations.

Observing the images mentioned above, it is possible to note that, with 50 iterations, the FSS+NBS algorithm, besides its low resolution, is capable of identifying the objects on the circular domain, unlike the other methods (having, however, an exception for the result of pure FSS, with the object placed on the edge of the interest's region). In 300 iterations, all methods are able to get images anatomically correct, considering the low resolution of EIT's images, being important to note that FSS and FSS+NBS showed better results than GA for this number of iterations.

The algorithms' final results (in 500 iterations) are of anatomically consistent and conclusive images, presenting little noise and good resolution. These results allow to conclude that the final obtained images of the three methods are similar in a qualitative analysis. Being so, the non-blind search application to FSS algorithm made little difference on the anatomical quality of the image, instead happened in previous works made by our research group, where the non-blind search algorithm, applied to Genetic Algorithm (Ribeiro et al., 2014c) and Particle Swarm Optimization (Feitosa et al., 2014) on the EIT problem, gave better results.

*Figure 2. Results using FSS for an object placed in the center (a1, a2 and a3), between the center and the edge (b1, b2 and b3) and on the edge (c1, c2 and c3) of the circular domain for 50, 300 and 500 iterations.*

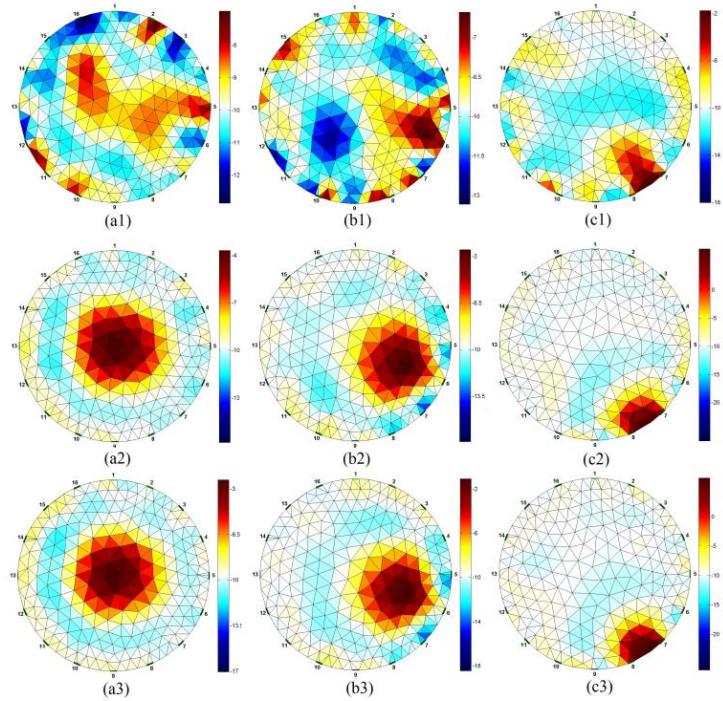

*Figure 3. Results using fish school search with non-blind search for an object placed in the center (a1, a2 and a3), between the center and the edge (b1, b2 and b3) and on the edge (c1, c2 and c3) of the circular domain for 50, 300 and 500 iterations.*

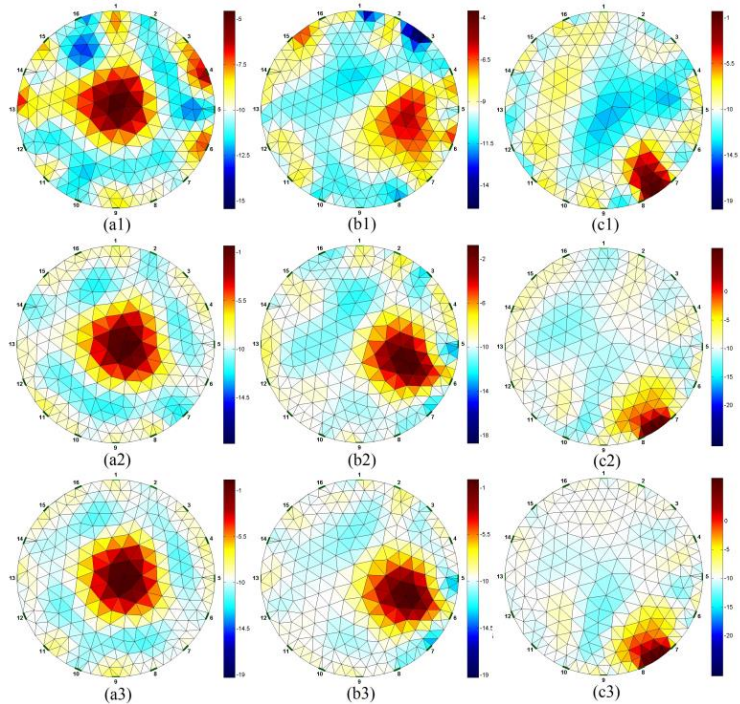

*Figure 4. Results using genetic algorithm for an object placed in the center (a1, a2 and a3), between the center and the edge (b1, b2 and b3) and on the edge (c1, c2 and c3) of the circular domain for 50, 300 and 500 iterations.*

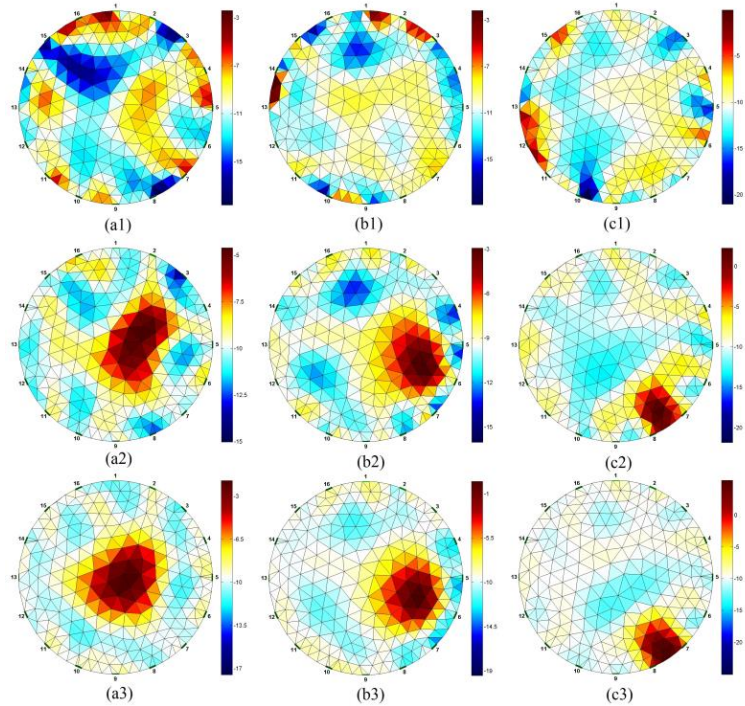

   Quantitatively, the algorithm's performance can be evaluated through the medium relative error (i.e., the fitness function) versus the iterations number graphic. The Figures 5, 6 and 7 show the 20 simulations average error decay versus the iterations number for the three reconstruction image's methods, in the situations where the object is placed at the center, between the center and the edge, and on the edge, respectively. Through these graphics, it is possible to observe that the behavior of these algorithms convergence is similar to an exponential decay, and, in the qualitative analysis, the mentioned decay of the objective function for FSS and FSS+NBS is quite similar to GA algorithm, having as an exception the fact that in the first iterations, this fall for the formers algorithms is more accentuated than the latter one. These quantitative results corroborate what was first noticed for the reconstructed images, in the qualitative analysis: FSS and FSS+NBS were able to generate consistent images with 300 iterations, although images obtained after 500 iterations were very similar for the three methods.

   During the experiments' execution, it was noticed a very high dependence of the Fish School Search algorithm relied on individual movement. In fact, this movement affects the fish weight update, the collective-instinctive and collective-volitive movements. Thus, the more well-succeeded fishes are those ones that presents bigger variation values of the fitness function (considering a minimization problem). This also explains the fact that the insertion of a fish on the first population, generated by another method as Non-Blind Search did not significantly enhance the algorithm's performance.

*Figure 5. Average error of 20 simulations in function of the number of iterations for the object in the center of the domain using Fish School Search without (FSS) and with Non-Blind Search (FSS+NBS) and Genetic Algorithm (AG).*

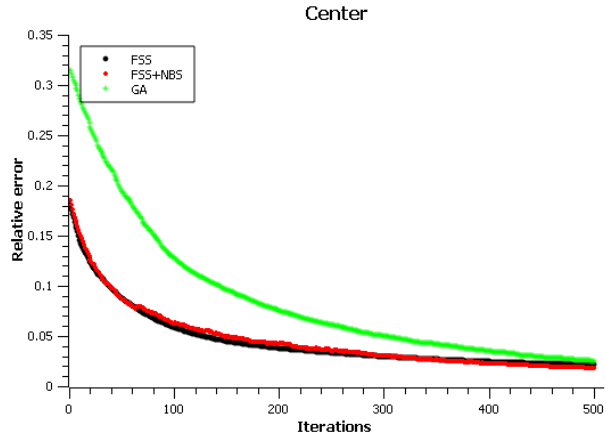

*Figure 6. Average error of 20 simulations in function of the number of iterations for the object between the center and the edge of the domain using Fish School Search without (FSS) and with Non-Blind Search (FSS+NBS) and Genetic Algorithm (AG).*

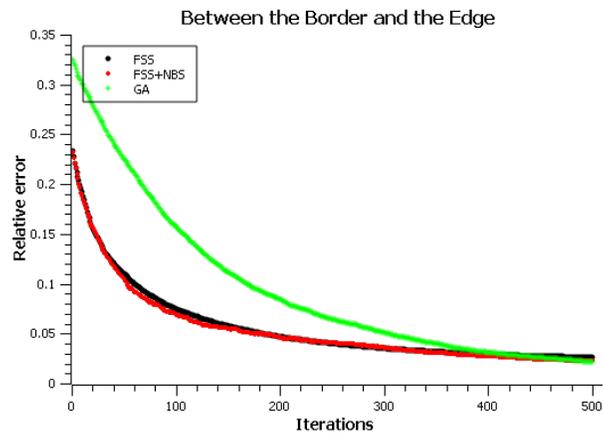

*Figure 7. Average error of 20 simulations in function of the number of iterations for the object on the edge of the domain using Fish School Search without (FSS) and with Non-Blind Search (FSS+NBS) and Genetic Algorithm (AG).*

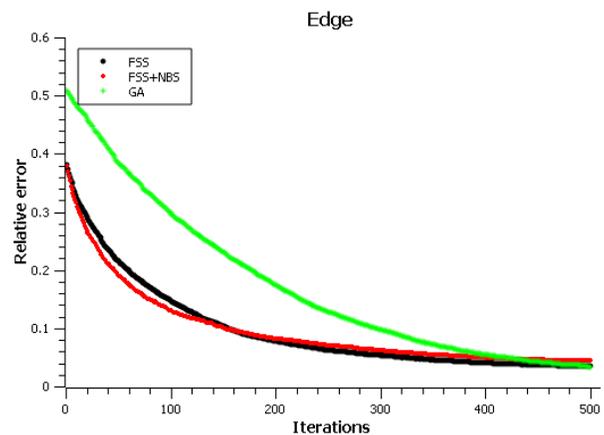

## CONCLUSION

Electric impedance tomography is a promising imaging technique, that has applications on engineering, sciences and medical sciences fields. Nowadays, the technique still presents low resolution images, which explains the researchers' efforts in this area. This work proposed and investigated the Fish School Search algorithm with and without Non-Blind Search on EIT images' reconstruction. In a general perspective, we can conclude that the use of fish school search with solution candidates obtained by using non-blind search based on Saha and Bandyopadhyay's Criterion (Saha & Bandyopadhyay, 2008) presented low contribution to the quantitative and qualitative FSS algorithm's performance, fact that can be explained by the algorithm's behavior, that favors the solution candidates considering its fitness variation during the iterative process and not the fitness value itself.

The obtained results for the here proposed methods for EIT problem's solution were compared to Genetic Algorithm, where the quantitative and qualitative results confirmed that Fish School Search is capable of generating results as good as the ones given by Genetic Algorithm.

For future works, looking forward to solve problems related to software, we propose the investigation of FSS' algorithm's hybridization with other methods, in order to improve EIT's image reconstruction, and to compare it with other methods in the actual Evolutionary Computing state of Art, including the hybridization with NBS. This research group will also focus on the migration of EIDORS from Matlab/Octave to a compiled or, at least, precompiled language, that supports experiments with parallel techniques and architecture, investigating software infrastructure and programming languages to achieve this goal.

From the hardware point of view, parallel architectures will be investigated, such as GPUs and clusters as and parallelism techniques, all of them to reduce the execution time of those algorithms. Evolutionary algorithms tend to load in its definitions a high parallelism level, and, with FSS, this situation is not different, explaining why it is important to invest in researches on the fields mentioned above.

## ACKNOWLEDGMENTS

The authors are grateful to the Brazilian scientific agencies CAPES and FACEPE, for the partial financial support of this work.